\documentclass[reprint, floatfix, pra, nofootinbib]{revtex4-1}

\usepackage{graphicx}
\usepackage{bm}
\usepackage{physics}
\usepackage{multirow}
\usepackage[table,dvipsnames,svgnames]{xcolor}
\usepackage[linkcolor=cyan]{hyperref}
\usepackage{svg}
\usepackage{amsmath}
\usepackage{amsfonts}
\usepackage{float}

\begin{document}

\title{String order melting of spin-1 particle chains in superconducting transmons using optimal control}
\author{Paul Kairys}
\email{pkairys@vols.utk.edu}
\affiliation{Bredesen Center for Interdisciplinary Research and Graduate Education, University of Tennessee, Knoxville, Tennessee}
\affiliation{Quantum Science Center, Oak Ridge National Laboratory, Oak Ridge, Tennessee}
\date{March 2021}
\thanks{This manuscript has been authored by UT-Battelle, LLC, under contract DE-AC05-00OR22725 with the US Department of Energy (DOE). The US government retains and the publisher, by accepting the article for publication, acknowledges that the US government retains a nonexclusive, paid-up, irrevocable, worldwide license to publish or reproduce the published form of this manuscript, or allow others to do so, for US government purposes. DOE will provide public access to these results of federally sponsored research in accordance with the DOE Public Access Plan (http://energy.gov/downloads/doe-public-access-plan).}

\author{Travis S.~Humble}
\email{humblets@ornl.gov}
\affiliation{Bredesen Center for Interdisciplinary Research and Graduate Education, University of Tennessee, Knoxville, Tennessee}
\affiliation{Quantum Science Center, Oak Ridge National Laboratory, Oak Ridge, Tennessee}
\date{March 2021}

\begin{abstract}
   Utilizing optimal control to simulate a model Hamiltonian is an emerging strategy that leverages the intrinsic physics of a device with digital quantum simulation methods. Here we evaluate optimal control for probing the non-equilibrium properties of symmetry-protected topological (SPT) states simulated with superconducting hardware. Assuming a tunable transmon architecture, we cast evolution of these SPT states as a series of one- and two-site pulse optimization problems that are solved in the presence of leakage constraints. From the generated pulses, we numerical simulate time-dependent melting of the perturbed SPT string order across a six-site model with an average state infidelity of $10^{-3}$. The feasibility of these pulses as well as their efficient application indicate that high-fidelity simulations of string-order melting are within reach of current quantum computing systems.
\end{abstract}

\maketitle

\section{Introduction}

Understanding the static and dynamical properties of quantum states is of paramount interest to the physical sciences. One route to enable these studies uses quantum devices and quantum information processing in protocols known as quantum simulation \cite{georgescu_quantum_2014}. There are several unique approaches to quantum simulation; from purely digital approaches on universal quantum computers to purely analog approaches using tailor-made quantum devices. Recent work suggests that there are also a number of intermediate paradigms capable of realizing quantum simulation \cite{salathe_digital_2015,parra-rodriguez_digital-analog_2020,lamata_digital-analog_2018,martin_digital-analog_2020,celeri_digital-analog_2021,lamata_digital-analog_2017,arrazola_digital-analog_2016,galicia_enhanced_2020,babukhin_hybrid_2020, kairys_parametrized_2021}. 
\par 
Quantum simulation based on quantum optimal control (QOC) permits one to take advantage of the natural device dynamics and Hilbert space while also enabling the use of digital decomposition methods~\cite{kairys_parametrized_2021}. In principle, this permits efficient use of coherent resources within quantum hardware. One of the most common applications of QOC is to identify the device controls which realize a desired unitary evolution \cite{glaser_training_2015}. Unitary evolution can then be used to preform state preparation, dynamical evolution, or even mitigate errors to improve information extraction \cite{glaser_training_2015}. 
\par
As a leading case study for quantum simulation, topological physics provide a route for both scientific discovery and engineering as well as the validation of quantum devices and simulation protocols.
One of the primary drivers of interest in these exotic phases of matter are their potential use within quantum information processing as robust quantum memories \cite{fowler_surface_2012,terhal2015quantum}. Some of these topological phases are well understood via both analytical and numerical methods that may be used for validating quantum simulation while others represent key open problems for the field \cite{terhal2015quantum}.
\par
Consider the example of one-dimensional symmetry-protected topological (SPT) phases, which exhibit interesting static and dynamical properties that are still under much study \cite{senthil_symmetry-protected_2015}. One of these phases, known as the Haldane phase, is realized in chains of interacting spin-1 particles that can be characterized by phenomena such as edge modes, degenerate entanglement spectra, and dilute (or hidden) antiferromagnetic order \cite{affleck_rigorous_1987,pollmann_entanglement_2010,pollmann_detection_2012,calvanese_strinati_destruction_2016}. A number of studies have shown that the latter phenomena, also called string order, can undergo a dynamic process called melting that leads to infinitesimally fast vanishing of string order under symmetry-breaking quantum quenches \cite{calvanese_strinati_destruction_2016}.
\par
Here we test the feasibility of studying string order melting using quantum simulation based on QOC. Our approach considers the simulation performed on superconducting quantum devices composed of coupled transmons operated as three-level (qutrit) systems. Transmon technology has matured significant over the past decade and such devices are currently available in configurations of up to $100$ transmons. Notably, these hardware systems are driven by analog control pulses that make them well suited for quantum simulation \cite{arute_quantum_2019,blok2021quantum,mckay2018qiskit}. In modeling the physics of these superconducting transmon devices, we demonstrate that control optimization can determine the local unitary evolutions that generate a symmetry-breaking quantum quench. These results are validated using  exact numerical simulation which confirm that feasibility of simulating string order melting in superconducting quantum devices.
\par
The remainder is outlined as follows: In Section~\ref{sec:proposed_exeriment}, the phenomena of string order melting is defined and strategies to observe the phenomena are discussed, based on Ref.~\cite{calvanese_strinati_destruction_2016}. In Section~\ref{sec:som_simulation}, we outline our  quantum simulation protocol of the quench dynamics and string order measurement. In Section~\ref{sec:device_architecture}, the device architecture is introduced and aspects of controlling this architecture are discussed. We report a set of optimal controls for this device architecture in Section~\ref{sec:optimal_control_results} and we validate that these controls enable the study of string order melting in Section~\ref{sec:validation_results}. Finally, we conclude and discuss avenues for future experimental demonstrations in Section~\ref{sec:conclusion}.

\section{String order melting}\label{sec:proposed_exeriment}


Studying dynamical many-body quantum systems is a classically challenging task because it requires integration of the Schr\"odinger equation on a Hilbert space which grows exponentially with increasing particle number. This difficulty motivates quantum simulation protocols to be used when studying such phenomena \cite{georgescu_quantum_2014}. In this work we propose a quantum simulation protocol to study string order melting analogous to the classical simulations performed by Calvanese et al. in Ref.~\cite{calvanese_strinati_destruction_2016} via classical numerical methods.
\par
The simulation of string-order melting requires evolving a quantum state with string order under symmetry-breaking time evolutions and observing the dynamics of string order as a function of string length, time, and direction. One of the best known models with string order is given by the AKLT model, named for Affleck, Kennedy, Lieb, and Tasaki. The AKLT Hamiltonian is defined as \cite{affleck_rigorous_1987}
\begin{align}\label{eq:aklt_ham}
    \hat{H}_{AKLT} &= \sum_{i=1}^N P(s_i + s_{i+1}=2) = \sum_{i=1}^N \mathcal{P}_{i,i+1},
\end{align}
where $N$ is the number of spin-1 particles, $s_i$ is the total spin of the particle on site $i$, and $P(s_i + s_{i+1}=2)=\mathcal{P}_{i,i+1}$ is the projector onto the subspace between two particles $i$ and $i+1$ with total spin equal to two: $s_i+s_{i+1}=2$.
\par
It was shown that this Hamiltonian has a unique set of ground states corresponding to the the mutual eigenvectors with eigenvalue $0$ for all the projectors $\mathcal{P}_{i,i+1}$. These states are known as valence bond solids and are naturally defined by projecting pairs of spin-1/2 particles in a singlet state into the spin-1 triplet subspace formed between singlets \cite{affleck_rigorous_1987,affleck1988valence}. The ground states of the AKLT Hamiltonian are contained within a phase called the Haldane phase which is a symmetry-protected topological phase with string order preserved by certain symmetries \cite{affleck_rigorous_1987,calvanese_strinati_destruction_2016,pollmann_entanglement_2010}. 
\par
To define string order one first defines an operator
\begin{align}\label{eq:string_operator}
    \hat{\mathcal{O}}^\alpha_{k,l} := \hat{S}_k^\alpha \bigg[ \prod_{n=k+1}^{l-1} e^{i \pi \hat{S}_n^\alpha}\bigg] \hat{S}_l^\alpha
\end{align}
where indices $k,n,l$ are lattice sites and $\hat{S}^\alpha_i$ is the spin-1 operator acting on lattice site $i$ which generates rotations around the $\alpha=x,y,z$ axes. The expectation value of $\hat{\mathcal{O}}^\alpha_{k,l}$ can be used to define an order parameter known as a string order parameter:
\begin{align}\label{eq:sop}
    \mathcal{O}^\alpha_{\text{string}}(\psi) = \lim_{|l-k| \rightarrow \infty} \bra{\psi} \hat{\mathcal{O}}^\alpha_{k,l} \ket{\psi}.
\end{align}
When $\mathcal{O}^\alpha_{\text{string}}(\psi) \neq 0$ the system is said to have string order and this order parameter is used to quantify string order melting. As defined in Ref.~\cite{calvanese_strinati_destruction_2016}, string order melting refers to the the decay of string order in the long-range limit of $|l-k| \rightarrow \infty$ at infinitesimal times. This phenomena is in stark contrast to typical Landau theory order parameters which cannot vanish instantly because of the continuity of time evolution \cite{calvanese_strinati_destruction_2016}. Thus string order melting represents a fascinating class of dynamical quantum physics in which quantum simulators may be particularly useful in studying.
\par
Following Ref.~\cite{calvanese_strinati_destruction_2016}, we consider a Hamiltonian under which the AKLT model is quenched to be a modified version of the spin-1 XXZ Hamiltonian:
\begin{align}\label{eq:quench_ham}
    \nonumber \hat{H}(\lambda, b) &= \sum_{i=1}^N \bigg[ \hat{S}^x_i \hat{S}^x_{i+1}+\hat{S}^y_i \hat{S}^y_{i+1} + \lambda \hat{S}^z_i \hat{S}^z_{i+1} \bigg] \\
     &+  b \sum_{i=1}^N \hat{S}^x_i
\end{align}
where $\lambda,b$ are two competing energy scales to be parameterized. It is known that for $b=0$ and $\lambda \leq \lambda_c \approx 1.186$ the ground state of the Hamiltonian is within the Haldane phase, the same phase as the AKLT ground states \cite{ueda2008finite}. The string order found within the Haldane phase is known to be preserved under perturbations invariant to the action of a symmetry group \cite{pollmann_entanglement_2010}
\begin{equation}
    \mathcal{G} = \{ \hat{1}, e^{i \pi \sum_i \hat{S}^x_i}, e^{i \pi \sum_i \hat{S}^y_i}, e^{i \pi \sum_i \hat{S}^z_i} \}. 
\end{equation}
This symmetry group is a specific representation of the dihedral group $D_2$ and any evolution which is invariant under the action of all elements of $\mathcal{G}$ will preserve string order. The transverse field term $b \sum_{i=1}^N \hat{S}^x_i$ in Eq.~(\ref{eq:quench_ham}) is not invariant under the action of all elements of $\mathcal{G}$ and therefore will lead to the loss of string order. However, in this case, the perturbation is invariant under the action of subsets of $\mathcal{G}$, which therefore preserves string order in the $x$ direction \cite{calvanese_strinati_destruction_2016}. The perturbation thus leads to melting of string order in the $z$ and $y$ directions only. We consider the implementation of quantum evolution under Eq.~(\ref{eq:quench_ham}).
\par
The simulation proceeds by initially preparing a ground state of the AKLT model on an open chain, evolving the state under Eq.~(\ref{eq:quench_ham}) with $\lambda = 0.2$ for a maximum time $T=2.5$ (here $\hbar=1$). At discrete time steps spaced by $\delta t = 0.1$, the string order observables $\hat{\mathcal{O}}^\alpha_{k,l}$ are calculated and the dependence on the length of the operator $l-k$, the direction $\alpha$, and the transverse field perturbation, $b$, are probed. These parameters are precisely the ones we will use to demonstrate that string order melting can be observed by quantum simulation enabled with optimal control however, other types of symmetry-breaking quenches could be explored in the future \cite{pollmann_entanglement_2010,calvanese_strinati_destruction_2016}.
\par
We compare the observed simulation results against a numerical simulation using exact diagonalization for $N=6$ spin-1 sites. From a physical point of view, this small size does not allow one to observe the large-$N$ effects which define string order melting. However, it does provide validation that the optimal controls found in this work enable quantitative simulation of the phenomena associated with string order melting.

\section{Quantum simulation of string order melting}\label{sec:som_simulation}

We consider generating dynamics of the Hamiltonian in Eq.~(\ref{eq:quench_ham}) with the transverse field strength parameter given as $b = \Delta b n_x$, where $\Delta b$ is the interval of $b$ on which we wish to explore quench dynamics and $n_x$ is an integer that determines the total magnitude of $b$. This re-parametrization allows us to decompose the global time evolution of the quantum quench into a product of local evolutions via Trotterization \cite{kairys_parametrized_2021}. In this work, the time evolution operator can be defined as
\begin{align}
 \nonumber U(\lambda, b,T_s) &= \exp\bigg(-\frac{iT_s}{\hbar} H(\lambda, b) \bigg) \\
 &= \lim_{q \rightarrow \infty} \bigg( \prod_{i=1}^{N}  U_{XYZ}^{i,i+1}\bigg(\frac{T_s}{q}\bigg) U_{X}\bigg(\frac{T_s}{q}\bigg)^{n_x}  \bigg)^q \label{eq:quench_trott},
\end{align}
where the $XYZ$ unitary determined by $\lambda$ is given by
\begin{equation}\label{eq:XYZ_trotter}
 U_{XYZ}^{i,i+1}\bigg(\frac{T_s}{q}\bigg) = \exp\bigg(-\frac{iT_s}{q\hbar} \bigg[ \hat{S}^x_i \hat{S}^x_{i+1}+\hat{S}^y_i \hat{S}^y_{i+1} + \lambda \hat{S}^z_i \hat{S}^z_{i+1}\bigg]\bigg)
\end{equation}
and the $X$ field unitary is given by
\begin{align}\label{eq:transverse_field_trotter}
 U_{X} \bigg(\frac{T_s}{q}\bigg) &= \exp\bigg(-\frac{iT_s \Delta b}{q\hbar} \sum_{i=0}^N \hat{S}^x_i \bigg) \\
 &= \prod_{i=0}^N \exp\bigg(-\frac{iT_s \Delta b}{q\hbar} \hat{S}^x_i \bigg). 
\end{align}
For the numerical demonstrations below, we consider $\Delta b=0.2$ and $\lambda=0.2$. 
\par
By truncating the limit in Eq.~(\ref{eq:quench_trott}) one obtains a $q$th-order approximation to the global evolution operator. Alternatively, we choose to use another common definition of the Trotter decomposition order given by a step size $\tau = T_s /q$. In this work we use optimal control methods to determine a set of device controls which generate the individual, local, Trotter evolutions. Then, by composing the local optimal controls in sequence, one is able approximate to the desired global quantum dynamics.
\par
We now show how to measure the string observables. We first introduce our notation for a system of $N$ spin-1 particles and demonstrate how measurements of the expectation value of the string-order operator can be evaluated. Consider the spin-1 operators $\hat{S}^\alpha$ with directional component $\alpha=x,y,z$. These operators obey the angular momentum commutation relations $[\hat{S}^\alpha,\hat{S}^\beta] = i\hbar \varepsilon_{\alpha \beta \gamma} \hat{S}^\gamma$, where $\varepsilon_{\alpha \beta \gamma}$ is the Levi-Civita symbol. The eigenvectors of each local operator are given by $\hat{S}^\alpha \ket{s^\alpha} = s\ket{s^\alpha}$ where $s=0,$ or $\pm$ labels the vectors with eigenvalues $0$ and $\pm 1$, respectively. One can convert from eigenstates of $\hat{S}^\beta$ to eigenstates of $\hat{S}^\alpha$ via the unitary operator $\hat{u}^{\alpha,\beta}$ defined as:
\begin{equation}
    \hat{u}^{\alpha,\beta} = \ket{-^\alpha} \bra{- ^\beta} + \ket{0^\alpha} \bra{0^\beta} + \ket{+^\alpha} \bra{+^\beta}.
\end{equation}
A tensor product of eigenstates with the same component $\alpha$ can be labeled via a string $\mathbf{s}$ as $\ket{\mathbf{s}^\alpha} = \otimes_{i=1}^N \ket{s_i^\alpha}$, where $s_i^\alpha$ is the $i$th element of the string $\mathbf{s}^\alpha$. Here $\ket{\mathbf{s}^\alpha}$ is a state in the composite Hilbert space of $N$ spin-1 particles and the set of all states formed by all possible strings $\mathbf{s}$ form a complete orthonormal basis for the composite Hilbert space, i.e., $\hat{I} = \sum_{\mathbf{s}} \ket{\mathbf{s}^\alpha}\bra{\mathbf{s}^\alpha}$. Moreover, one can transform between tensor product basis states using $\hat{U}^{\alpha,\beta} = \bigotimes_{i=1}^N\hat{u}^{\alpha,\beta}$:
\begin{equation}
    \ket{\mathbf{s}^\alpha} = \hat{U}^{\alpha,\beta}\ket{\mathbf{s}^\beta}=\bigotimes_{i=1}^N \hat{u}^{\alpha,\beta}\ket{\mathbf{s}^\beta}
\end{equation}
We will now consider the expansion of an expectation value of a string order operator upon a particular direction $\alpha$ for an arbitrary state $\ket{\psi}$ (See Appendix~\ref{sec:string_order_parameter_appendix} for full derivation)
\begin{align}
    \bra{\psi} \hat{\mathcal{O}}^\alpha_{k,l} \ket{\psi} &= \bra{\psi} \bigg( \hat{S}_k^\alpha \bigg[ \prod_{n=k+1}^{l-1} e^{i \pi \hat{S}_n^\alpha}\bigg] \hat{S}_l^\alpha \bigg) \ket{\psi} \\
    &= \sum_\mathbf{s} \bigg(s_k \bigg[ \prod_{n=k+1}^{l-1} e^{i \pi s_n}\bigg] s_l \bigg) \bigg|\bra{\mathbf{s}^z} \hat{U}^{\alpha z}\ket{\psi}\bigg|^2 \label{eq:string_expectation}
\end{align}
where we have $\hat{U}^{z,\alpha}$ to convert from the $\hat{S}^\alpha$ basis to the $\hat{S}^z$ basis. This decomposition shows that by applying $\hat{U}^{z,\alpha}$ after preparing the state $\ket{\psi}$ permits measurements in the standard $\hat{S}^z$ basis that are the same as measuring in the  $\hat{S}^\alpha$ basis. Then, when we observe a measurement outcome string $\mathbf{s}$ we can calculate the weights in the sums of Eq.~(\ref{eq:string_expectation}). Once we have the weights and the probabilities of measuring a particular string, we can easily estimate $\bra{\psi} \hat{\mathcal{O}}^\alpha_{k,l} \ket{\psi}$ on a classical computer. Therefore, to estimate the string order operator expectation value on a device requires implementing $\hat{U}^{xz}$ and $\hat{U}^{yz}$. We define $\hat{U}^{xz}$ and $\hat{U}^{yz}$ in the standard $\hat{S}_z$ basis in Eqs.~(\ref{eq:zx_transform}) and (\ref{eq:zy_transform}), respectively.

\section{Device architecture and problem mapping}\label{sec:device_architecture}


There are a variety of routes to realize quantum simulation of spin-1 systems. One route is to construct tailored quantum devices with spin-1 degrees of freedom, such as cold atoms, trapped ions, or strongly correlated superconducting circuits \cite{yip2003dimer,garcia2004implementation,imambekov2003spin,hilker2017revealing,cohen2014proposal,cohen2015simulating,senko2015realization,albarran2018spin}. However such devices have either not been constructed or their construction and control may be infeasible with current technology. Development and engineering of such novel quantum devices would require significant, and potentially unknown, time and cost.
\par
An alternative route is to adapt currently developed quantum hardware to perform the desired simulation, leveraging a tremendous body of work in the understanding and engineering of such devices. There are a number of existing device paradigms which can enable quantum simulation of spin-1 systems via qubit-based digital quantum computing. Unfortunately, there is a distinct overhead in using qubits to model spin-1 systems that will be particularly inconvenient for near- and mid-term realizations. This overhead arises because the local Hilbert space of each spin-1 particle is of dimension three, requiring at least two qubits to represent each spin-1 particle. 
\par
To eliminate this overhead it is therefore natural to consider quantum devices in which the basic physical element has local Hilbert space of at least three. One of the leading device paradigms which can satisfy this requirement are superconducting devices based on transmons \cite{krantz_quantum_2019}. Each transmon is a nonlinear oscillator and computations can be performed in the low energy subspace of these systems, allowing one to create, in principle, systems of qudits \cite{krantz_quantum_2019,wu_high-fidelity_2020}. Recently, systems of multiple interacting transmons operated as qutrits have been demonstrated, suggesting a feasible route for enabling quantum simulations of interacting spin-1 systems \cite{blok2021quantum}.
\par
We choose a mapping of the device-model Hilbert spaces that identifies the eigenstates of the local spin-1 $z$ operator $\hat{S}^z$ ($\ket{-},\ket{0},\ket{+}$) with the eigenstates of the local excitation number operator $\hat{n}$ ($\ket{0},\ket{1},\ket{2}$) of each transmon:
\begin{align}\label{eq:hilb_mapping}
    \ket{-} &\rightarrow \ket{0}\\ \nonumber
    \ket{0} &\rightarrow \ket{1}\\ \nonumber
    \ket{+} &\rightarrow \ket{2}.
\end{align}
This composite Hilbert space formed by a system of $N$ spin-1 particles can be realized by a system of $N$ transmons. For all discussions that follow this mapping will be used. 
\par
Having selected the basic quantum information element, we now state our assumptions about the device architecture. A variety of superconducting device architectures based on transmons have been developed \cite{krantz_quantum_2019} including multiple types of transmons and modalities of transmon interactions. We consider an architecture of tunable-frequency transmons mediated by tunable couplers, as explored previously s~\cite{li2022realization,yan_tunable_2018,sung_realization_2020}. Similar architectures have been generalized to large devices with more than 50 transmons operating at high fidelity \cite{arute_quantum_2019}. 
\par
Within this architecture, transmons are modeled as coupled anharmonic (Duffing) oscillators with interactions between transmons mediated by a tunable coupler \cite{krantz_quantum_2019,li2022realization,yan_tunable_2018,sung_realization_2020}. The effective device Hamiltonian then becomes
\begin{align}\label{eq:duffing_hamiltonian}
    \nonumber H &= \sum_{i=1}^N \bigg[ \omega_{i} \hat{n}_i + \frac{\delta_i}{2} \hat{n}_i(\hat{n}_i - 1) \\
    \nonumber &~~~~~~~+ \delta\omega_{i}(t) \hat{n}_i+ \varepsilon_i(t) (\hat{a}^\dagger_i + \hat{a}_i) \bigg] \\
    &~~~~~~~+ \sum_{\langle i,j \rangle} g_{i,j}(t)(\hat{a}^\dagger_i \hat{a}_{j} + \hat{a}_i \hat{a}^\dagger_{j})
\end{align}
where the operators $\hat{a}_i^\dagger,\hat{a}_i$ are Bosonic creation and annihilation operators, respectively, and $\hat{n}_i = \hat{a}^\dagger_i \hat{a}_i$ is the Bosonic number operator. The parameters $\omega_i,\delta_i$ are the idling frequency and anharmonicity of transmon $i$. The time-dependent functions in Eq.~(\ref{eq:duffing_hamiltonian}), are the frequency detunings $\delta\omega_i(t)$ of each transmon, the local microwave controls are given by $\varepsilon_i(t)$, and the tunable coupling is given by $g_{i,j}(t)$. 
\begin{figure*}
    \centering
    \includegraphics[width=1.0\textwidth]{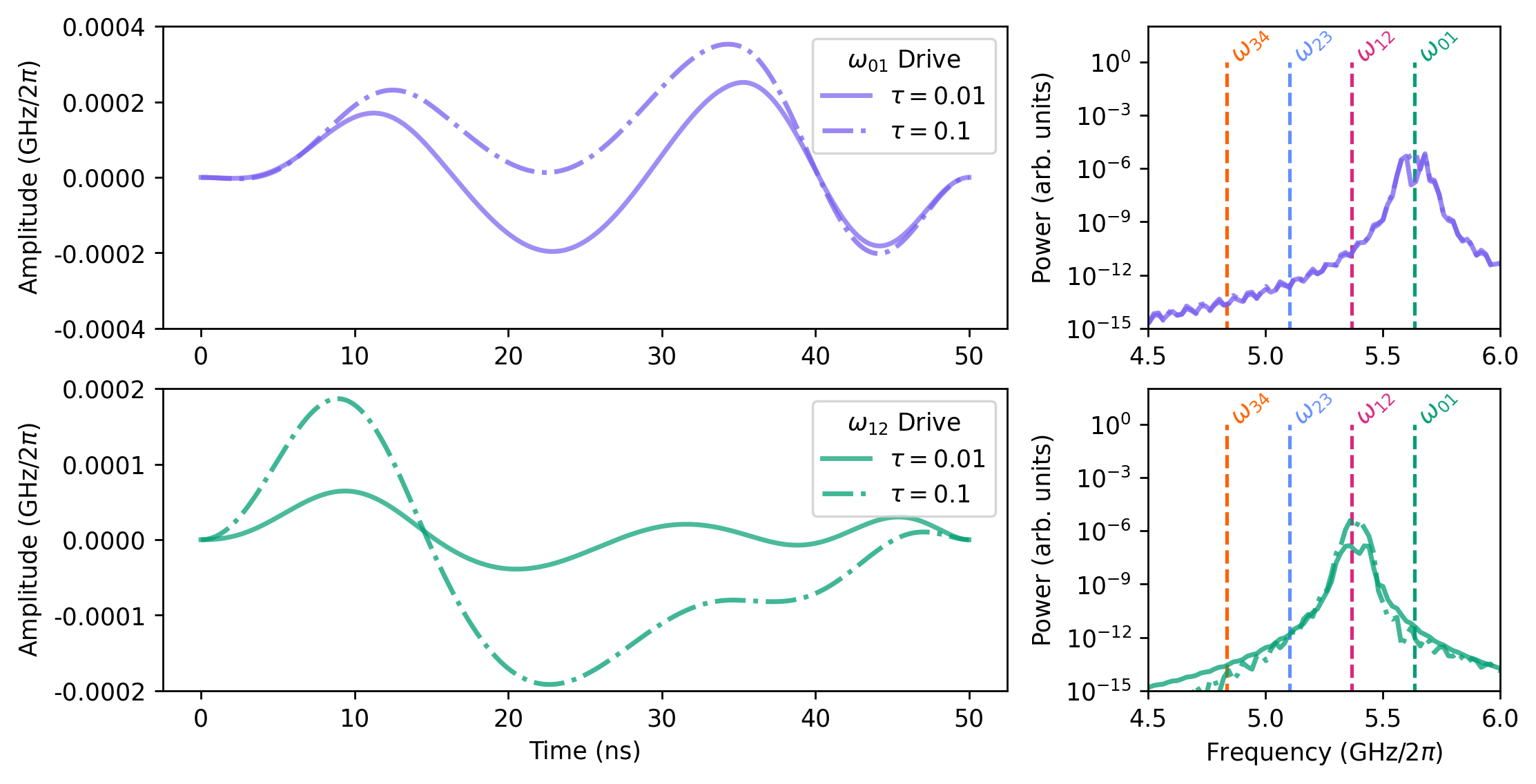}
    \caption{Optimal microwave controls for a single transmon that generates the single-site Trotter evolution for the quench dynamics for two different Trotter-step sizes $\tau=0.1$ and $\tau=0.01$. (Top panels) The optimal pulse envelopes found via optimal control modulated at the transmon $0\rightarrow 1$ frequency $\omega_{01}$. The frequency spectrum of these pulses in the lab frame are shown to the right of the time-domain pulses, with key transition frequencies labeled. (Bottom panels) The optimal pulse envelopes modulated at the transmon $1 \rightarrow 2$ transition frequency $\omega_{12}$. The frequency spectrum of these pulses in the lab frame are shown to the right of the time-domain pulses. The solid lines and dashed-dot lines refer to the optimal pulses that generate the Trotter-step unitary Eq.~(\ref{eq:transverse_field_trotter}) at $\tau=0.1$ and $\tau=0.01$, respectively.}
    \label{fig:control_results_1}
\end{figure*}
\begin{figure*}
    \centering
    \includegraphics[width=1.0\textwidth]{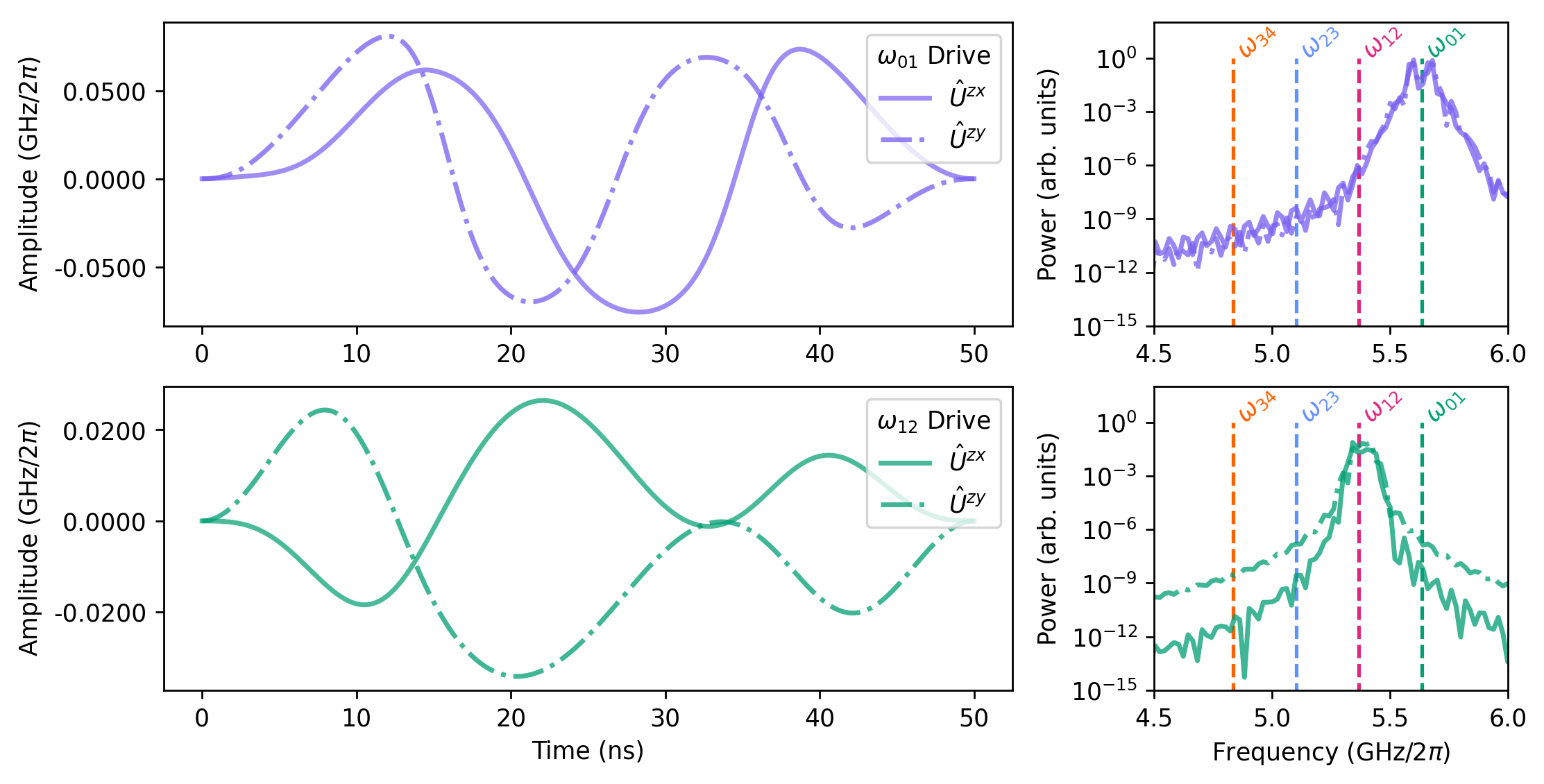}
    \caption{Optimal microwave controls for a single transmon that generates the rotation operators $\hat{U}^{zx}$ and $\hat{U}^{zy}$, enabling measurement of the string order parameter in the assumed device architecture. (Top panels) The optimal pulse envelopes found via optimal control modulated at the transmon $0\rightarrow 1$ frequency $\omega_{01}$. The frequency spectrum of these pulses in the lab frame are shown to the right of the time-domain pulses, with key transition frequencies labeled. (Bottom panels) The optimal pulse envelopes modulated at the transmon $1 \rightarrow 2$ transition frequency $\omega_{12}$. The frequency spectrum of these pulses in the lab frame are shown to the right of the time-domain pulses. The solid lines and dashed-dot lines refer to the optimal pulses that generate the unitaries Eq.~(\ref{eq:zx_transform}) and Eq.~(\ref{eq:zy_transform}).}
    \label{fig:control_results_2}
\end{figure*}
The density and layout of the transmons in a real system depends on a number of practical limitations such as calibration complexity and noise \cite{arute_quantum_2019}. since we are only considering a one-dimensional chain of spin-1 particles, it suffices to assume that each transmon is connected to at most two neighboring transmons in order to form the needed one dimensional topology. This ensures that the unitaries  implemented via optimal controls are local within the device. Additional connections to ancilla transmons may be useful for state preparation or observable measurements but we do not consider these possibilities here.
\par
\par
We consider a general task of embedding a spin-$s$ particle into the first $2s+1$ levels of an oscillator. This can be done compactly by defining a map between the eigenvalues $n$ of the excitation number operator $\hat{n}$ and the eigenvalues of the spin-$s$ $z$ operator $\hat{S}_z$: $\hat{S}_z\ket{n} = (n+s)\ket{n} $ for $n\leq (2s+1)$, where $\hat{n}\ket{n} = n\ket{n}$ are the eigenvectors of the local excitation number operator. 
\par
We define the $z$ component of the total-spin operator for $M$ spins as $\hat{J}_z = \sum_{i=1}^M S_z^{(i)}$, which satisfies the eigenvalue equation $\hat{J}_z\ket{m}=m\ket{m}$. Now, we consider the action of $\hat{J}_z$ on a state with a fixed excitation number $\ket{\psi} = \otimes_{i=1}^M \ket{n_i}$ with $n_i \leq (2s+1)$:
\begin{align}
    \hat{J}_z\ket{\psi} &= \otimes_{i=1}^M \hat{S}_z\ket{n_i} \\
    &= \otimes_{i=1}^M (n_i+s)\ket{n_i} \\
    &= \bigg[\prod_{i=1}^M (n_i+s)\bigg] \ket{\psi}.
\end{align}
In other words, the $z$ component of total spin of state $\ket{\psi}$ is $\prod_{i=1}^M (n_i+s)$. Most importantly, we note that the eigenvalue of $\hat{J}_z$ will be the same for every state with the same number of excitations because the eigenvalue $m = \prod_{i=1}^M (n_i+s)$ is invariant to permutations of the number of excitations between sites.
\par
Intuitively, this means that all states with a fixed $z$ component of total spin $m$ lie within the subspace spanned by states with a fixed particle number. This is important when considering implementing evolutions which will preserve the $z$ component of total spin, which are common. For example, typical two-local interactions of the form $\hat{S}_+^{(i)}\hat{S}_-^{(j)}+\hat{S}_+^{(i)}\hat{S}_-^{(j)} \propto S_x^{(i)}S_x^{(j)}+S_y^{(i)}S_y^{(j)}$ preserve the $z$ component of total spin (as do Heisenberg-type interactions), and, therefore, any action generated by such terms would be constrained to operate within a fixed particle-number block of the oscillator system. 
\par
These considerations limit the dynamics for the two-site Trotter step defined in Eq.~(\ref{eq:XYZ_trotter}) to a block-diagonal representation of the total particle number basis for the transmon system. Therefore, analog device controls which preserve total particle number are the most natural controls to use to attempt to generate the unitary defined in Eq.~(\ref{eq:XYZ_trotter}). 
\par
Examining the assumed device Hamiltonian reveals that all the terms in the Hamiltonian except for the microwave control will preserve particle number. Thus we  generate the desired two-site Trotter step unitary, Eq.~(\ref{eq:XYZ_trotter}), using only transmon frequency and coupling controls. This choice of device controls reduces simulation complexity and potentially reduces the complexity of pulse calibration/characterization because the dynamics during gate operation are constrained to blocks of total particle number.

\section{Optimal control results}\label{sec:optimal_control_results}
\begin{figure*}
    \centering
    \includegraphics[width=1.0\textwidth]{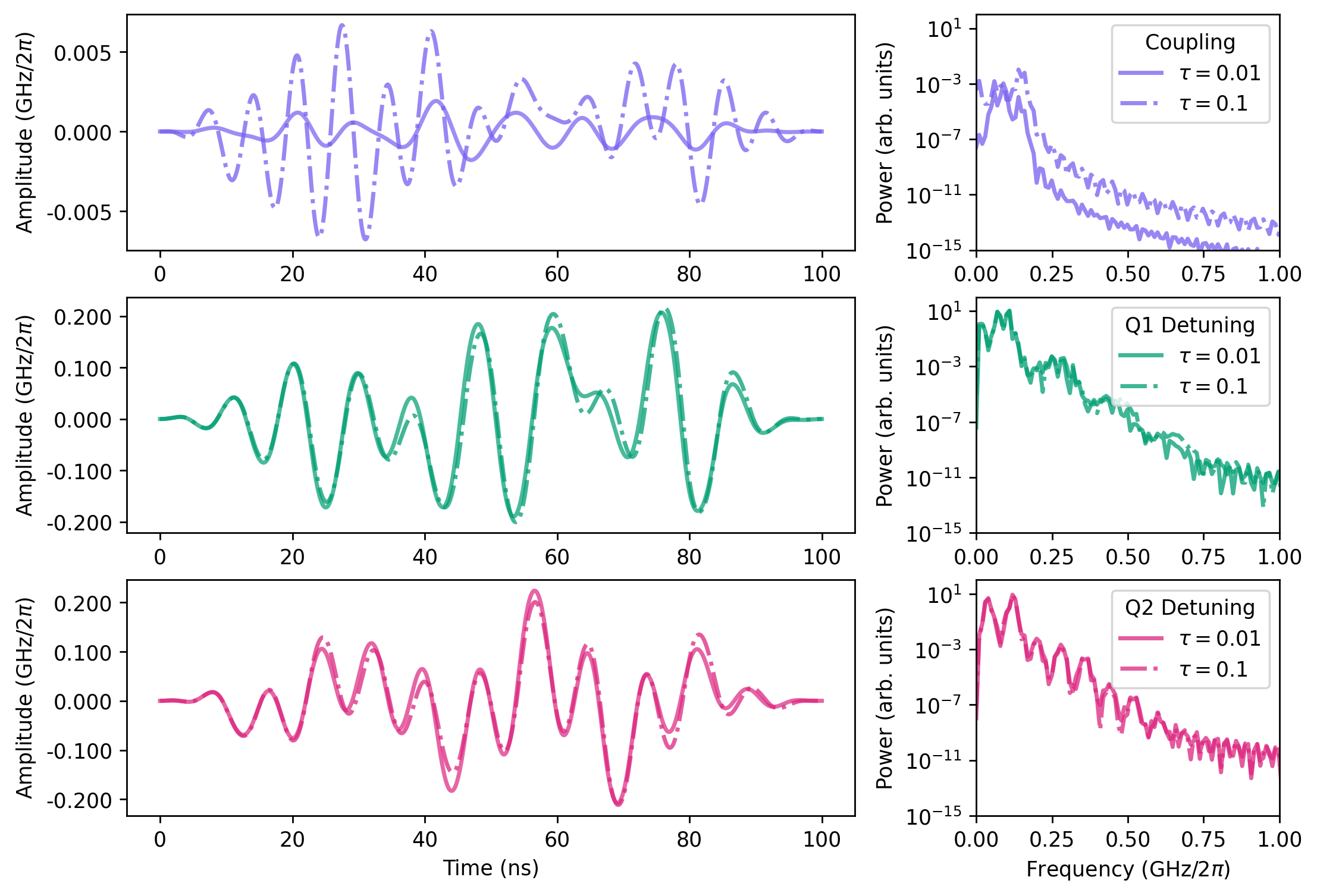}
    \caption{Optimal controls for a two-transmon system that generates the two-site Trotter evolution for the quench dynamics (Eq.~(\ref{eq:XYZ_trotter})) for two different Trotter-step sizes $\tau=0.1$ and $\tau=0.01$. (Top panels) The optimal coupling controls found via optimal. The frequency spectrum of these pulses in the lab frame are shown to the right of the time-domain pulses. (Middle panels) The optimal detuning controls for transmon $1$ found via optimal. The frequency spectrum of these pulses in the lab frame are shown to the right of the time-domain pulses. (Bottom panels) The optimal detuning controls for transmon $2$ found via optimal. The frequency spectrum of these pulses in the lab frame are shown to the right of the time-domain pulses. The solid lines and dashed-dot lines refer to the optimal pulses that generate the Trotter-step unitary Eq.~(\ref{eq:XYZ_trotter}) at $\tau=0.1$ and $\tau=0.01$, respectively.}
    \label{fig:control_results_4}
\end{figure*}
\begin{figure*}
    \centering
    \includegraphics[width=1.0\textwidth]{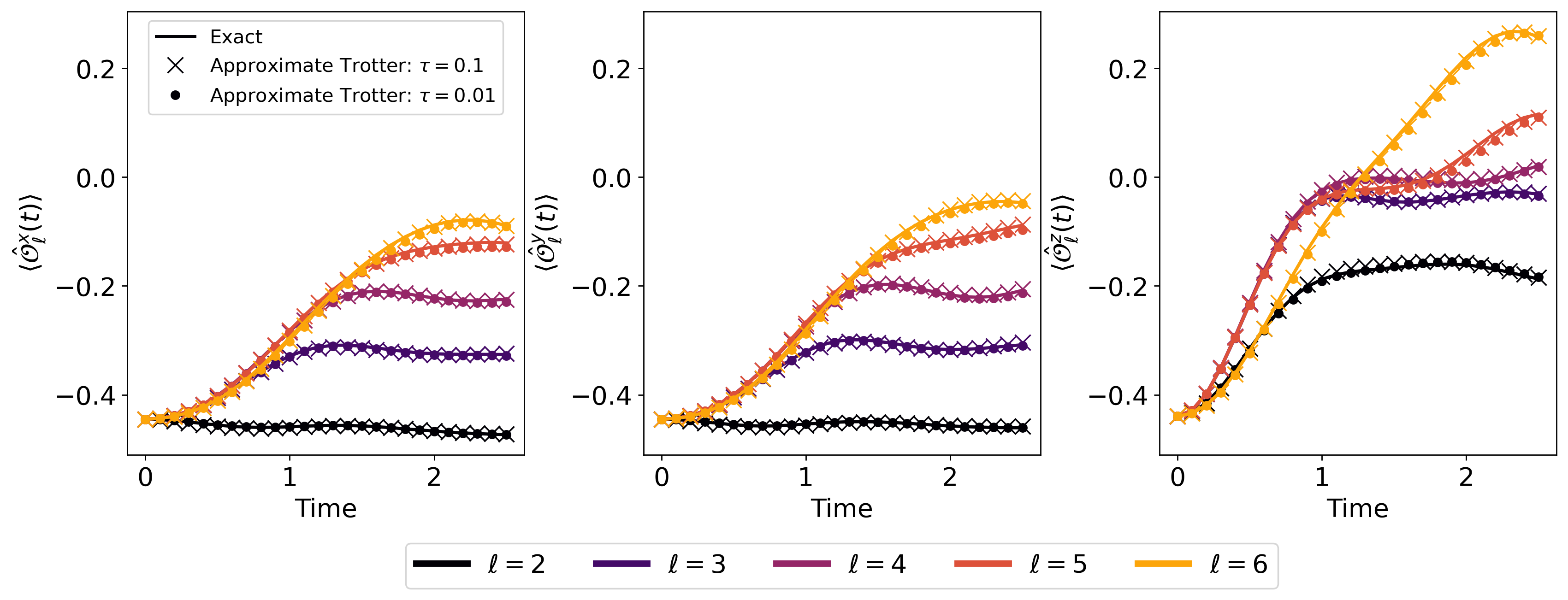}
    \caption{Time dynamics of the string observable in different directions $\alpha = x,y,~z$ (left to right) for a 6-site AKLT model as a function of string length when $b = 0.2$. The solid lines are exact dynamics generated by numerical diagonalization of the global 6-site quench Hamiltonian, Eq.~(\ref{eq:quench_ham}). The markers represent the approximate Trotterized dynamics generated via repeated application of the final-time unitaries identified via optimal control for two different Trotter step sizes $\tau=0.1$ and $\tau=0.01$.}
    \label{fig:dynamic_results_1}
\end{figure*}
In Section~\ref{sec:som_simulation}, we outlined a set of four unitary operations to simulate and observe the phenomena of string order melting within a superconducting transmon device. The first two sets of unitaries, Eqs.~(\ref{eq:XYZ_trotter}) and (\ref{eq:transverse_field_trotter}), simulate the time dynamics of string order melting. The second two, $\hat{U}^{zx}$(Eq.~(\ref{eq:zx_transform})) and $\hat{U}^{zy}$(Eq.~(\ref{eq:zy_transform})), rotate into the correct basis so that string order information can be extracted from the device. 
\par
In this section, we analyze a set of optimized device controls capable of generating these unitaries. We begin by analyzing the unitaries defined on single transmons: Eqs.~(\ref{eq:transverse_field_trotter}),~(\ref{eq:zx_transform}), and (\ref{eq:zy_transform}) and then proceed to discuss optimal controls for a two-transmon system which generate Eq.~(\ref{eq:XYZ_trotter}). The numerical methods used to generate these controls are detailed in Appendix~\ref{sec:methods}.
\par
The Trotter-step unitary for a single-site transverse field perturbation is defined in Eq.~(\ref{eq:transverse_field_trotter}) and can be generated by modulating microwave-frequency controls at the transition frequency of the transmon. In Fig.~\ref{fig:control_results_1}, we show a set of optimal pulses at the two main transmon frequencies, $\omega_{01}$ and $\omega_{12}$. These plots compare optimal controls for Trotter-step sizes $\tau=0.1$ and $\tau=0.01$ with infidelities (as defined in Eq.~(\ref{eq:infidelity})) of $\approx 5\times10^{-10}$ and $\approx 1\times10^{-10}$, respectively. 
\par
From the results shown in Fig.~\ref{fig:control_results_1}, we find that the optimal control pulses for a $50$ ns control time have high fidelity for two key reasons: 1) the pulses are well localized in the frequency domain which prevents transitioning population to higher levels and 2) the amplitude of the control pulses are very low, which suppresses the off-resonant excitations. 
\par
The next results are for the basis changes needed to measure string order in the transmon system: $\hat{U}^{zx}$(Eq.~(\ref{eq:zx_transform})) and $\hat{U}^{zy}$(Eq.~(\ref{eq:zy_transform})). The optimal microwave controls which generate both unitaries were defined to be $50$ ns and are shown in Fig.~\ref{fig:control_results_2}. In both cases, we observe similar pulse characteristics such as amplitude limits and shifts in the frequency domain around the primary drive frequencies. The primary difference between the pulses is the small phase offset of the $\omega_{01}$ drive and large phase offset of the $\omega_{12}$ drive. The phase of the pulses that generate $\hat{U}^{zy}$ are offset from those that generate $\hat{U}^{zx}$ because the phase of the microwave control pulses selects the axis of rotation in the $x-y$ plane. Another slight difference is the infidelity of the pulses, which is $\approx 3\times10^{-6}$ and $\approx 2\times10^{-5}$ for $\hat{U}^{zx}$ and $\hat{U}^{zy}$, respectively. 
\par
The difference in infidelity is due to different amounts of residual leakage outside of the computational subspace. It can be seen in the power spectrum of the optimal $\omega_{12}$ drive for $\hat{U}^{zy}$ has about an order of magnitude more amplitude on the $\omega_{23}$ transition frequency, contributing more to leakage. Why the optimizer was unable to find a better solution is most likely due to the presence of a local minima in the optimization landscape induced by our pulse constraints \cite{riviello_searching_2015} and would probably be resolved by more exhaustive numerical searches. However, we do not perform such exhaustive search because these pulses need only be applied once after the state evolution and therefore achieving significantly higher fidelity only negligibly affects the overall simulation accuracy. 
\par
For the two-transmon optimal controls, in Fig.~\ref{fig:control_results_4} we show a set of optimal controls for Trotter-step sizes $\tau=0.1$ and $\tau=0.01$ with infidelities of $\approx 4\times10^{-6}$ and $\approx 4\times10^{-8}$, respectively. The similarity between the optimal pulses is primarily in the transmon detuning (lower two rows) because the optimal controls for $\tau=0.1$ were used as the starting point for the optimization of $\tau=0.01$. However, there is a striking difference between the optimal coupling controls in the average amplitude of the coupling, which corresponds to $0$~GHz in the frequency domain. This decrease can be empirically understood as leading to less generated rotation around the desired axis during the control time.
\par
Overall, the optimal controls we identified meet a number of criteria necessary for implementation in real devices. All of the pulses we found are well within experimental bandwidth limitations and the control amplitudes are also within feasible limits \cite{krantz_quantum_2019,sung_realization_2020,li2022realization}. Moreover, each control pulse requires about $60$ parameters to describe via the control ansatz we used (defined in Appendix~\ref{sec:methods}); perhaps fewer if a more optimal basis is chosen. This suggests that implementing and calibrating these pulses should be feasible in modern superconducting devices \cite{egger_adaptive_2014,arute_quantum_2019,sung_realization_2020,li2022realization}. 


\section{Validation of melting dynamics with optimal control}\label{sec:validation_results}
\begin{figure*}
    \centering
    \includegraphics[width=1.0\textwidth]{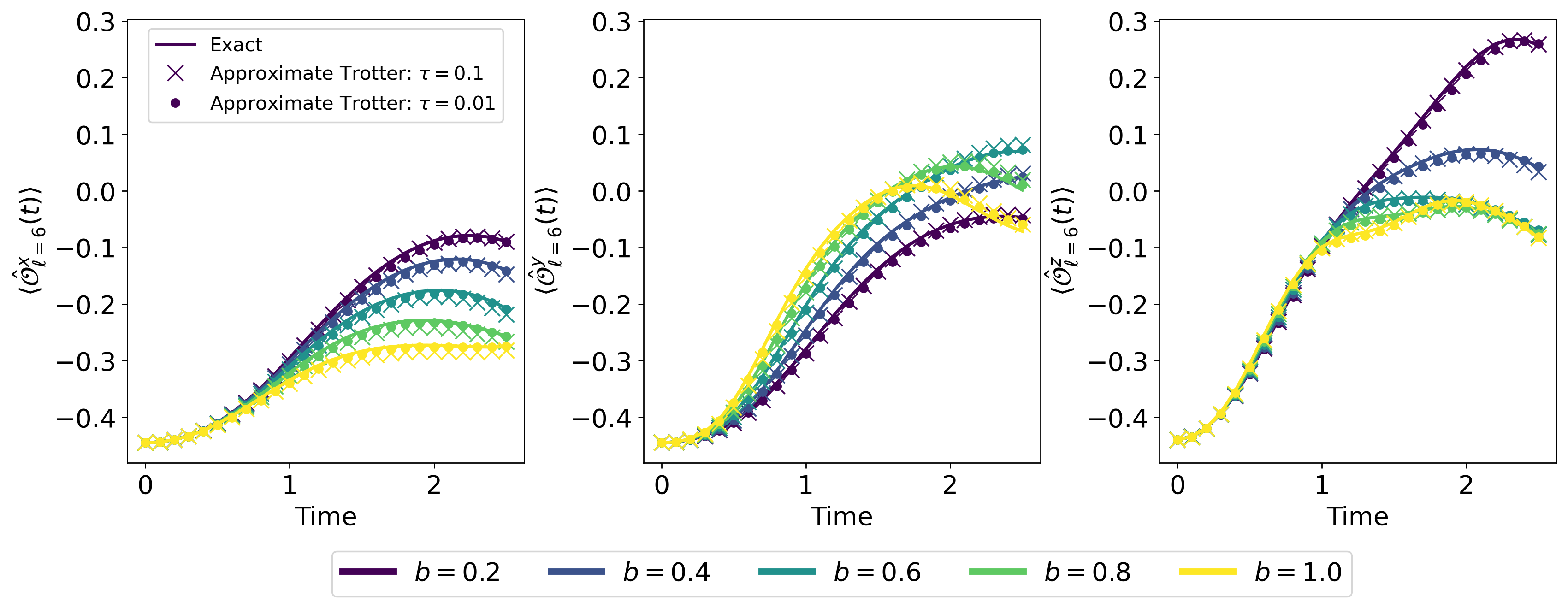}
    \caption{Time dynamics of the $l=6$ string observable in different directions $\alpha = x,y,~z$ (left to right) for a 6-site AKLT model as a function of perturbation strength $b$. The solid lines are exact dynamics generated by numerical diagonalization of the global 6-site quench Hamiltonian, Eq.~(\ref{eq:quench_ham}). The markers represent the approximate Trotterized dynamics generated via repeated application of the final-time unitaries identified via optimal control for two different Trotter step sizes $\tau=0.1$ and $\tau=0.01$. The single-transmon optimal controls were intended to generate a perturbation of $b=0.2$, thus to realize $b = 0.2n$ the single-transmon optimal controls are applied $n$ times per Trotter layer.}
    \label{fig:dynamic_results_2}
\end{figure*}
\begin{figure}
    \centering
    \includegraphics[width=1.0\columnwidth]{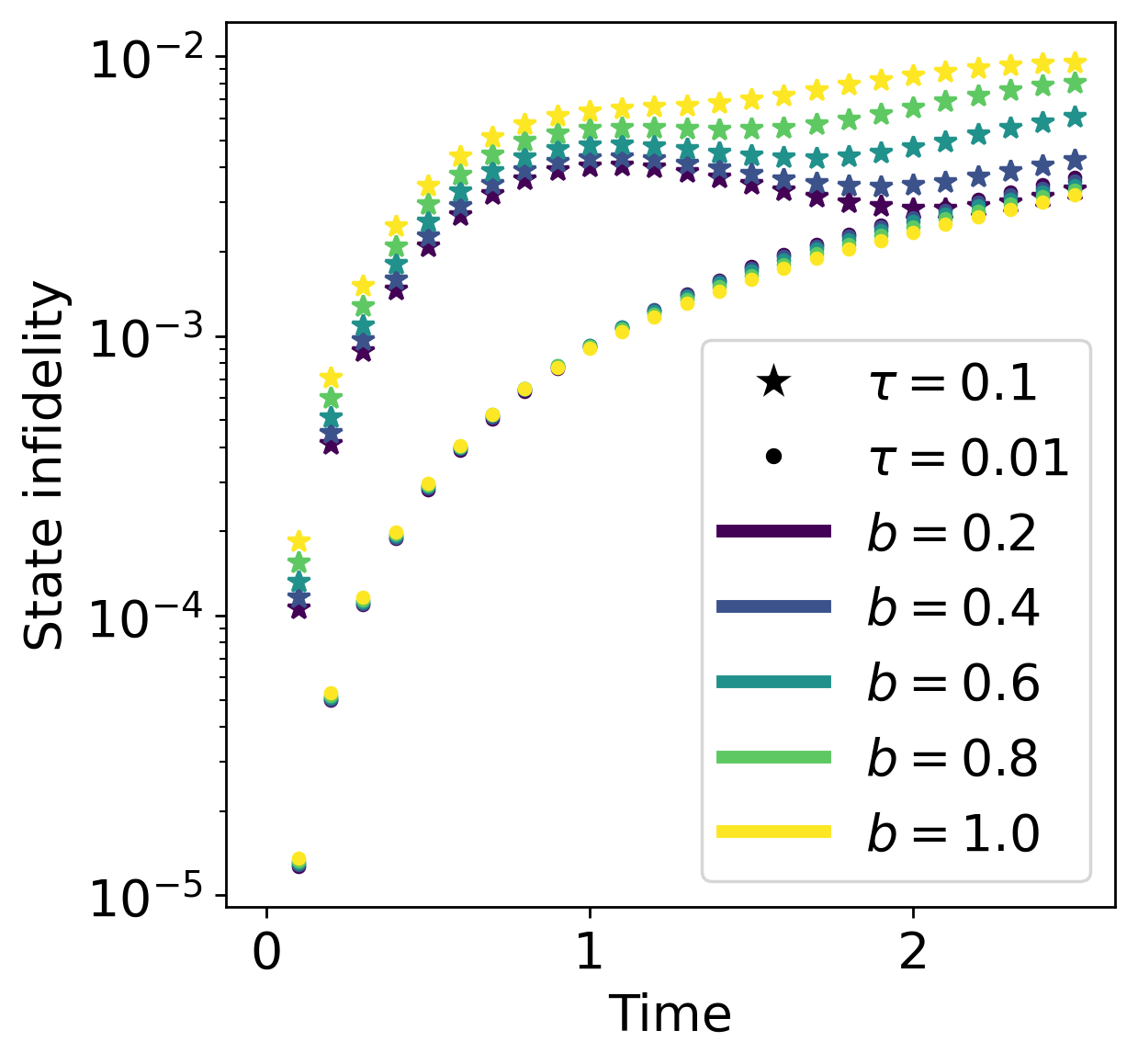}
    \caption{Time dynamics of state infidelity for a 6-site AKLT state generated by the Trotterized quench dynamics found via optimal control for increasing strengths of the symmetry-breaking perturbation. The two different marker types represent dynamics approximating the true dynamics at two different Trotter step sizes $\tau=0.1$ and $\tau=0.01$.}
    \label{fig:dynamic_results_3}
\end{figure}
We now present results from numerical simulations to validate that the optimal controls identified enable the simulation of string order melting. We first extract the unitaries generated by the optimal controls in the previous section. For computational convenience, we choose to make a Markovian assumption about the device dynamics, where we neglect population generated outside of the computational subspace. This leads to a non-unitary simulation but we find that the errors introduced by our optimal controls are less than those induced by Trotter error. 
\par
First, in Fig.~\ref{fig:dynamic_results_1} we plot the time dynamics of the string observables for each spatial direction as a function of time and string length for a weak transverse field perturbation $b=0.2$. We observe that there is an initial decrease in the string order parameter in every direction between $t=0$ and $t=1$ which is in agreement with Ref.~\cite{calvanese_strinati_destruction_2016} for short strings. Importantly, we observe that the Trotter step of $\tau=0.1$ is capable of tracking the qualitative dynamics for each observable with only limited quantitative error.
\par
Next, in Fig.~\ref{fig:dynamic_results_2}, we plot the dynamics of the $l=6$ string order operator in each spatial direction. We observe that for increasing strength of the perturbation increases the rate of string order destruction for $\langle \hat{\mathcal{O}}^y (t)\rangle$ and $\langle \hat{\mathcal{O}}^z (t)\rangle$ but leads to less destruction of string order for $\langle \hat{\mathcal{O}}^x (t)\rangle$. This is understood because the perturbation being applied only partially breaks the protecting symmetry and no string order melting occurs in the $x$ direction. This is, again in qualitative agreement with the observations made in Ref.~\cite{calvanese_strinati_destruction_2016} for small string lengths.
\par
Finally, in Fig.~\ref{fig:dynamic_results_3} we show the state fidelity as a function of perturbation and for the  Trotterized dynamics generated by the optimal controls. As expected, we see that the smaller Trotter-step size yields a lower state fidelity and we observe that the infidelity is much less dependent on the perturbation strength for $\tau=0.01$. Importantly, we observe a non-monotonic change in state fidelity for the larger Trotter step size and a monotonic one for $\tau=0.01$. This indicates that choosing an optimal Trotter step size for this problem may be difficult, depending on the timescale of dynamics one wishes to observe. Because string order melting is a fast dynamical phenomena a smaller $\tau$ may be better because the accuracy of the simulation as determined by state fidelity seems to change logarithmically at short times.

\section{Conclusion}\label{sec:conclusion}

In this work we have presented a path towards quantum simulation of string order dynamics in a superconducting transmon architecture. We have used quantum optimal control to verify that the necessary quantum processes can be implemented and our numerical simulations based on these optimal controls indicate that experimental observation of string order melting should be feasible in the a near generation of superconducting devices. 
Our approach to designing these simulations uses superconducting transmons to simulate spin-1 particles, microwave control lines to generate excitations, and tunable couplers and tunable frequency transmons to drive particle-conserving evolution that generates entanglement between transmons. 
\par
We employed numerical optimization methods to generate device controls that are capable of driving the designed unitary evolutions with high fidelity. The characteristics of these controls indicate that optimal pulses could be implemented within existing devices, even when accounting for known hardware limitations. We have used these controls to simulate the time dynamics of the AKLT state in the presence of a quantum quench, and we have verified that these simulations predict a number of qualitative and quantitative markers of string order melting for SPT states.
\par
The analyses of these simulations have further quantified the dependence of state fidelity on evolution time, perturbation strength, and Trotter decomposition order. We found that a choosing smaller time step $\tau$ is significantly less sensitive to the perturbation strength and generally leads to a monotonic decrease in state fidelity. However, it is important to note that a smaller choice of $\tau$ requires a proportionally longer evolution time because the number of repeated applications of the Trotter step. In this case, $\tau=0.01$ requires a factor of 10 more gates to realize the same simulation time of $\tau=0.1$. In the presence of decoherence or control errors, the well known trade off in errors generated by Trotterization versus gate depth are expected to limit the total time simulated \cite{salathe_digital_2015,knee2015optimal}. 
\par
Our results support simulating the dynamics of string order melting in a transmon architecture as experimentally feasible, but there are still a number of challenges for implementation that require further consideration. The most immediate route of future work is through additional numerical simulations.
\par
A clearer understanding of the relationships between the duration of the controls, their infidelity, and ultimately their robustness to noise is needed. We have fixed the pulse duration for single (50 ns) and two-transmon (100 ns) optimal control simulations but optimal controls with comparable fidelities may exist at smaller control times \cite{caneva_optimal_2009,kirchhoff_optimized_2018}. Optimal pulses with smaller control times would enable the observation of longer time dynamics within the same device coherence times. Furthermore, quantifying or improving the robustness of controls against noise within the applied field will aide in their implementations \cite{egger_adaptive_2014,lysne2020small}. 
\par
Another challenge is state preparation of the entangled spin-1 states within the Haldane phase, such as the AKLT state. There are a variety of routes to achieve this that may be feasible for near-term devices, for example,  variational state preparation and adiabatic or digitized adiabatic evolution \cite{barends_digitized_2016,kandala_hardware-efficient_2017}. In particular, because the Haldane phase is symmetry-protected topological phase, it can be adiabatically connected to a product state via an evolution that breaks all of the symmetries that protect the phase and therefore can be efficiently prepared via adiabatic methods \cite{pollmann_entanglement_2010}.
\par
Another approach to state preparation that is less efficient, but perhaps complimentary, is to prepare the AKLT state by projecting into the ground-state subspace via a sequence of parity measurements. This can be achieved by noting that the AKLT Hamiltonian is a sum of Hermitian projectors that can be converted to involutory Hermitian operators. These projections can be implemented by controlled evolution using an ancilla qubit to perform a type of parity measurement that projects the state into a subspace where the AKLT state lies. This non-unitary route to preparing the AKLT state would also enable direct measurement of the terms within the Hamiltonian or even a route for dissipative state preparation \cite{kraus2008preparation,verstraete2009quantum}.
\par
In conclusion, we have shown how quantum simulation using quantum optimal control offers a unique approach to study the static and dynamical properties of a model Hamiltonian. Using numerical simulations, we have confirmed that the quantum simulation of dynamical topological phenomena such as string order melting are feasible in current superconducting devices. This affords many new opportunities for the implementation and optimization of quantum simulations. 

\section*{Acknowledgements}
This material is based upon work supported by the U.S.~Department of Energy, Office of Science, National Quantum Information Science Research Centers, Quantum Science Center and the U.S.~Department of Energy, Office of Science, Early Career Research Award. This research used resources of the Compute and Data Environment for Science (CADES) at the Oak Ridge National Laboratory, which is supported by the Office of Science of the U.S. Department of Energy under Contract No. DE-AC05-00OR22725.
\par
This manuscript has been authored by UT-Battelle, LLC, under Contract No. DE-AC05-00OR22725 with the U.S. Department of Energy (DOE). The U.S. Government retains and the publisher, by accepting the article for publication, acknowledges that the U.S. Government retains a nonexclusive, paid-up, irrevocable, worldwide license to publish or reproduce the published form of this manuscript, or allow others to do so, for U.S. Government purposes.

\appendix

\section{String order parameter measurement}\label{sec:string_order_parameter_appendix}

In this section we present the derivation of Eq.~(\ref{eq:string_expectation}) for the expectation value of a string order operator for an arbitrary state $\ket{\psi}$ in the direction $\alpha$:
\begin{widetext}
\begin{align}
    \bra{\psi} \hat{\mathcal{O}}^\alpha_{k,l} \ket{\psi} &= \bra{\psi} \bigg( \hat{S}_k^\alpha \bigg[ \prod_{n=k+1}^{l-1} e^{i \pi \hat{S}_n^\alpha}\bigg] \hat{S}_l^\alpha \bigg) \ket{\psi} 
    = \sum_\mathbf{s} \bra{\psi} \bigg( \hat{S}_k^\alpha \bigg[ \prod_{n=k+1}^{l-1} e^{i \pi \hat{S}_n^\alpha}\bigg] \hat{S}_l^\alpha \bigg)\ket{\mathbf{s}^\alpha} \bra{\mathbf{s}^\alpha}\ket{\psi} \\
    &= \sum_\mathbf{s} \bra{\psi} \bigg(s_k \bigg[ \prod_{n=k+1}^{l-1} e^{i \pi s_n}\bigg] s_l \bigg) \ket{\mathbf{s}^\alpha} \bra{\mathbf{s}^\alpha}\ket{\psi} 
    = \sum_\mathbf{s} \bigg(s_k \bigg[ \prod_{n=k+1}^{l-1} e^{i \pi s_n}\bigg] s_l \bigg)\bra{\psi}\ket{\mathbf{s}^\alpha} \bra{\mathbf{s}^\alpha}\ket{\psi} \\
    &= \sum_\mathbf{s} \bigg(s_k \bigg[ \prod_{n=k+1}^{l-1} e^{i \pi s_n}\bigg] s_l \bigg) \bra{\psi}U^{z \alpha \dagger} U^{z \alpha}\ket{\mathbf{s}^\alpha} \bra{\mathbf{s}^\alpha}U^{z \alpha \dagger} U^{z \alpha}\ket{\psi} \\
    &= \sum_\mathbf{s} \bigg(s_k \bigg[ \prod_{n=k+1}^{l-1} e^{i \pi s_n}\bigg] s_l \bigg) \bra{\psi}U^{\alpha z \dagger} \ket{\mathbf{s}^z} \bra{\mathbf{s}^z} U^{\alpha z}\ket{\psi}.
\end{align}
\end{widetext}
Using the definitions of the spin-1 operators and the commutation relations, we express the unitary operators $\hat{U}^{z,\alpha}$ in the basis of eigenvectors of $\hat{S}^z$:
\begin{align}\label{eq:zx_transform}
    \hat{U}^{zx} &= \frac{1}{2} \ket{-} \bigg( \bra{+} - \sqrt{2} \bra{0} + \bra{-} \bigg) \\ 
    \nonumber &+ \frac{1}{\sqrt{2}} \ket{0} \bigg( -\bra{+}+ \bra{-} \bigg) \\
    \nonumber &+ \frac{1}{2} \ket{+} \bigg( \bra{+} + \sqrt{2} \bra{0} + \bra{-} \bigg)
\end{align}
and
\begin{align}\label{eq:zy_transform}
    \hat{U}^{zy} &= \frac{1}{2} \ket{-} \bigg( -\bra{+} + i\sqrt{2} \bra{0} + \bra{-} \bigg) \\ 
    \nonumber &+ \frac{1}{\sqrt{2}} \ket{0} \bigg( \bra{+}+ \bra{-} \bigg) \\
    \nonumber &+ \frac{1}{2} \ket{+} \bigg( -\bra{+} -i \sqrt{2} \bra{0} + \bra{-} \bigg).
\end{align}
\section{Numerical methods}\label{sec:methods}

We utilize numerical optimal control techniques to simulate the dynamics generated by Eq.~(\ref{eq:duffing_hamiltonian}) and iteratively optimize the time-dependent device controls to minimize the infidelity between the evolved system dynamics and a target unitary operator. These methods and implementations follow what has been discussed in previous work \cite{kairys_parametrized_2021,kairys2021efficient} and we review the details for completeness and convenience. 
\par
We identify optimal controls that generate four unitary operators as defined in Eqs.~(\ref{eq:quench_trott}),~(\ref{eq:zx_transform}),~(\ref{eq:zy_transform}), and (\ref{eq:XYZ_trotter}). The first three unitaries are each defined on a single-transmon Hilbert space and therefore we use a single transmon Hamiltonian derived from Eq.~(\ref{eq:duffing_hamiltonian}) that omits the frequency detuning control and uses only microwave control parameters:
\begin{align}
    H_D &= \omega_{1} \hat{n}_1 + \frac{\delta_1}{2} \hat{n}_1(\hat{n}_1 - 1) +  \varepsilon_1(t) (\hat{a}^\dagger_1 + \hat{a}_1).
\end{align}
For the two-site Trotter step unitary, we use a two-transmon Hamiltonian derived from Eq.~(\ref{eq:duffing_hamiltonian}) omitting the microwave control lines because the target unitary, Eq.~(\ref{eq:XYZ_trotter}), is block-diagonal in the total particle number basis, such that
\begin{align}
    H_D &= \omega_{1} \hat{n}_1 + \frac{\delta_1}{2} \hat{n}_1(\hat{n}_1 - 1) + \delta\omega_{1}(t) \hat{n}_1 \\
    &+ \omega_{2} \hat{n}_2 + \frac{\delta_2}{2} \hat{n}_2(\hat{n}_2 - 1) + \delta\omega_{2}(t) \hat{n}_2 \\
    &+~g_{1,2}(t)(\hat{a}^\dagger_1 \hat{a}_{2} + \hat{a}_1 \hat{a}^\dagger_{2}).
\end{align}
In both cases, quantum information processing is performed in reference to a pre-calibrated rotating frame associated with the idling frequency of each transmon. This transformation is given by the unitary transform
\begin{align}\label{eq:rotating_frame}
 R(t) &=  \bigotimes_i R_i(t) \\
 &= \bigotimes_i \exp \bigg[ \frac{i t}{\hbar} \omega_{i} \hat{n}_i \bigg]
\end{align}
which yields an effective Hamiltonian in this rotating frame, $H^R = i\hbar (\partial_t R(t)) R^\dagger(t) + R(t)H_D R^\dagger(t)$, and the corresponding time-ordered evolution operator for control time $T_c$ of a transmon system is
\begin{equation}\label{eq:device_unitary}
 U_D(T_c) = \mathcal{T}\exp \bigg[ -\frac{i}{\hbar} \int_0^{T_c} d\tau H^{R}(\tau) \bigg].
\end{equation}
We then use the Gradient Optimization of Analytic Controls (GOAT) algorithm to evaluate gradients of the objective function with respect to control parameters \cite{machnes_tunable_2018}. Our primary objective function to be minimized is the unitary infidelity of a quantum processes using a projective $SU$ measure derived from the Hilbert-Schmidt inner product \cite{palao_optimal_2003}:
\begin{equation}\label{eq:infidelity}
 g(\vec{\alpha}) = 1 - \frac{1}{d^2} \bigg\vert Tr(U^\dagger_{T} P_{c}U_D(\vec{\alpha},T_c)P_{c}) \bigg\vert ^2
\end{equation}
where $U_T$ is the target unitary operator we wish to prepare, $U_D(\vec{\alpha},T_c)$ is the unitary evolution operator for the device with control parameters $\vec{\alpha}$ and control time $T_c$, $P_c$ is a projection onto the desired computational subspace and $d$ is the dimension of the computational subspace.
\par
Within the GOAT algorithm, each control field $f(t)$ (e.g., $\delta\omega_i(t)$, $\varepsilon_i(t)$, $g_{i,j}(t)$) is decomposed into a (not-necessarily orthonormal) function basis which is parameterized by a set of real numbers -- enabling traditional  numerical optimization techniques to be used to optimize the device controls. 
\par
Here we describe each control field via a functional form which includes both the control parameters to be optimized as well as additional functions which constrain the optimization to a class of practical pulses. For all of our optimizations we describe each control field $f(t)$ (e.g., $\delta\omega_i(t)$, $\varepsilon_i(t)$, $g_{i,j}(t)$) as
\begin{align}\label{eq:control_ansatz}
    f(\vec{\alpha},t) = \Omega(t)\cos(\overline{\omega}t)S(h(\vec{\alpha},t))
\end{align}
Where we have defined a carrier frequency $\overline{\omega}$, a window function $\Omega(t)$ to ensure that the pulse turns on and off smoothly, a saturation function $S(x)$ to ensure that the optimal pulses stay within a pre-specified amplitude range, and the parameterized function $h(\vec{\alpha},t)$. The window function is a flat-top cosine defined as 
\begin{align}
    \Omega(t) = 
    \begin{cases} 
      \frac{1-\cos(\pi t/\tau_r)}{2}\Omega_m & 0\leq t \leq \tau_r \\
      \Omega_m & \tau_r \leq t \leq (\tau_c-\tau_r) \\
      \frac{1-\cos(\pi (\tau_c -t)/\tau_r)}{2} \Omega_m & (\tau_c - \tau_r) \leq t \leq \tau_c,
   \end{cases}
\end{align}
where $\tau_c= T_c$ is the total control time, $\Omega_m$ scales the magnitude of the control pulse, and $\tau_r$ is the ramp time, which was chosen to be $0.3\tau_c$ to reduce spectral leakage \cite{tripathi_operation_2019}. The saturation function is a generalized logistic function defined as
\begin{equation}
    S(x) = -B - \frac{2B}{1-3\exp(-\frac{4 x}{B})}
\end{equation}
where $B=0.08\text{ GHz}/2\pi$ for microwave controls, $B=0.5\text{ GHz}/2\pi$ for frequency detuning controls, $B=0.01\text{ GHz}/2\pi$ for coupling controls. These parameters were chosen to agree roughly with control limitations currently observed in superconducting transmon devices of the assumed architecture \cite{sung_realization_2020,li2022realization}. Finally, we expand each control field as a linear combination of $N$ analytic functions, which we choose to be sinusoidal functions with varying amplitude, frequency, and phase:
\begin{equation}\label{eq:sinusoidal}
    h(\vec{\alpha},t) = \bigg[ \sum_{n}^N \alpha_{n,1}\sin(\alpha_{n,2}t+\alpha_{n,3})\bigg].
\end{equation}
For the microwave controls operating on a single transmon we optimize two drive channels as in Eq.~(\ref{eq:control_ansatz}) each at a different carrier frequency $\overline{\omega} = \omega_{01}, \omega_{12}$, which are the transition frequencies between the transmon levels $\ket{0}\rightarrow \ket{1}$ and $\ket{1}\rightarrow \ket{2}$, respectively. We also assume the control time for a single qutrit operation is $T_c=50$~ns. For each channel we set $N=10$ in Eq.~(\ref{eq:sinusoidal}). This means that each of the single-transmon control pulses require optimization of $60$ parameters.
\par
For the coupling control and frequency detuning controls we choose a carrier frequency of $\overline{\omega}=0$, which places the dynamics within those typically used to generate resonant two-qubit gates on superconducting hardware \cite{sung_realization_2020,krantz_quantum_2019}. We assume a control time of $100$~ns for the two-transmon evolution because this coincides with experimental timescales realizable in current devices \cite{sung_realization_2020,li2022realization}. Moreover, we set $N=7$ in Eq.~(\ref{eq:sinusoidal}) for each control channel which gives a total of $63$ total parameters for optimization.
\par
For the microwave controls we penalize controls with intermediate-time leakage to higher subspaces by computing a functional and adding it to the infidelity in Eq.~(\ref{eq:infidelity}):
\begin{align}\label{eq:leakage}
    \mathcal{L}(\vec{\alpha}) &= \frac{1}{T_c}\int_0^{T_c} d\tau \Tr( P_{c}U^\dagger(\vec{\alpha},\tau)P_{d}U(\vec{\alpha},\tau)P_{c})
\end{align}
where we have defined $P_c$ as the projector onto the computational subspace of the transmon (In our simulations of spin-1 particles, this means span$(\{\ket{0},\ket{1},\ket{2}\})$ and the operator $P_{d}$ assigns weights to specific leakage levels depending on the importance that little intermediate population lie in that state. In our simulations we set $P_d = 0.1 \ket{3}\bra{3}+ 1\ket{4}\bra{4}$ to weakly penalize leakage to $\ket{3}$ and strongly on $\ket{4}$.
\par
The precise choice of weights for the leakage subspaces was not significant as we constrained the optimizer to look for pulses well localized within the frequency domain at the allowed transition frequencies. However the leakage penalty does help ensure that the optimal pulses identified will generalize to a true anharmonic quantum oscillator with an infinite number of energy levels.
\par
In order to perform gradient based optimization we require the gradient of this function with respect to a parameter $\alpha$ we derive the gradient of the leakage penalty as
\begin{align}
    \partial_\alpha \mathcal{L} &= \partial_\alpha \frac{1}{T_c}\int_0^{T_c} d\tau \Tr( P_{c}U^\dagger(\vec{\alpha},\tau)P_{d}U(\vec{\alpha},\tau)P_{c}) \\
    &= \frac{1}{T_c}\int_0^{T_c} d\tau \partial_\alpha \Tr( P_{c}U^\dagger(\vec{\alpha},\tau)P_{d}U(\vec{\alpha},\tau)P_{c}) \\
    &= \frac{1}{T_c}\int_0^{T_c} d\tau \Tr \bigg( P_c \partial_\alpha U^\dagger(\vec{\alpha},\tau) P_d U(\vec{\alpha},\tau) P_c \\
    &~~~~~~~~~~~~~~~~~~~~+ P_c U^\dagger(\vec{\alpha},\tau) P_d \partial_\alpha U(\vec{\alpha},\tau) P_c \bigg)
\end{align}
and we see that calculating this gradient requires only knowledge of $\partial_\alpha U(\vec{\alpha},t)$, which is obtained through the GOAT method \cite{machnes_tunable_2018}. A leakage penalty was not added to the two-site operator optimization because non-computational states are actively being used for quantum information processing in that case \cite{kairys_parametrized_2021}. 
\par
In all simulations, we do not implement the rotating wave approximation to ensure a more accurate estimate of gate fidelity. Moreover, we model each transmon as a 5-level system to fully account for leakage in the two-transmon evolutions and to accurately quantify leakage in the single transmon simulations. 
\par
We perform optimizations to identify controls that generate the Trotter-step operators (Eqs.~(\ref{eq:XYZ_trotter}) and (\ref{eq:transverse_field_trotter})) at two Trotter step sizes: $\tau=0.1$ and $\tau = 0.01$. We seed the optimization for Trotter step size $\tau=0.1$ with a random initial guess and $\tau = 0.01$ with the optimal controls obtained for $\tau=0.1$. 
\par
We use experimental hardware parameters for the frequency and anharmonicity of each transmon derived from Ref.~\cite{blok2021quantum} in which a system of transmons are used as qutrits. Specifically we define $\omega_1 = 5.634\text{ GHz}/2\pi$, $\delta_1 = -0.266\text{ GHz}/2\pi$, $\omega_2 = 5.447\text{ GHz}/2\pi$, and $\delta_2 = -0.270\text{ GHz}/2\pi$. 
\par
We implement the GOAT algorithm using the programming language Julia and various open-source packages \cite{Bezanson_Julia_A_fresh_2017}. Our implementation uses the Julia package DifferentialEquations.jl to numerically solve the coupled GOAT equations of motion using a order 5/4 Runge-Kutta method with adaptive time stepping \cite{rackauckas2017differentialequations}. For the gradient-based control optimization of $\vec{\alpha}$, we use a limited-memory Broyden-Fletcher-Goldfarb-Shanno (L-BFGS) algorithm with a backtracking line-search method which are implemented in the Optim.jl package and LineSearches.jl package, respectively \cite{mogensen2018optim}. We limit each optimization to 2000 iterations of L-BFGS and define a stopping criteria when the infinity-norm of the gradient falls below 1e-9 or the relative change in the objective function is below 1e-8. For further details on the derivations of gradients via the GOAT algorithm we refer the reader to our previous work \cite{kairys_parametrized_2021}. 
\par
Finally, to validate that our resulting optimal controls should enable the observation of string order melting in real superconducting hardware we perform a numerical simulation to observe the dynamics of string order under the optimized controls. Specifically, we compute the dynamics of a 6-site spin-1 system using the resulting optimal controls and compare the results obtained via exact numerical integration of the model. We initialize the system in the AKLT state obtained via diagonalization of the AKLT Hamiltonian Eq.~(\ref{eq:aklt_ham}). Then, we evolve the state under the exact quench dynamics given by Eq.~(\ref{eq:quench_ham}) and compare these with the dynamics generated by two sets of Trotterized dynamics generated via optimal controls for two different Trotter step sizes: $\tau=0.1$ and $\tau = 0.01$.  We then compute the string order operator expectation value and state fidelity to draw our final conclusions.

\bibliographystyle{unsrt}
\bibliography{references}

\begin{thebibliography}{10}

\bibitem{georgescu_quantum_2014}
I.~M. Georgescu, S.~Ashhab, and F.~Nori.
\newblock Quantum simulation.
\newblock {\em Reviews of Modern Physics}, 86(1):153--185, March 2014.

\bibitem{salathe_digital_2015}
Y.~Salathé, M.~Mondal, M.~Oppliger, J.~Heinsoo, P.~Kurpiers, A.~Potočnik,
  A.~Mezzacapo, U.~Las~Heras, L.~Lamata, E.~Solano, S.~Filipp, and A.~Wallraff.
\newblock Digital {Quantum} {Simulation} of {Spin} {Models} with {Circuit}
  {Quantum} {Electrodynamics}.
\newblock {\em Physical Review X}, 5(2):021027, June 2015.

\bibitem{parra-rodriguez_digital-analog_2020}
A.~Parra-Rodriguez, P.~Lougovski, L.~Lamata, E.~Solano, and M.~Sanz.
\newblock Digital-analog quantum computation.
\newblock {\em Physical Review A}, 101(2):022305, February 2020.

\bibitem{lamata_digital-analog_2018}
Lucas Lamata, Adrian Parra-Rodriguez, Mikel Sanz, and Enrique Solano.
\newblock Digital-analog quantum simulations with superconducting circuits.
\newblock {\em Advances in Physics: X}, 3(1):1457981, January 2018.

\bibitem{martin_digital-analog_2020}
A.~Martin, L.~Lamata, E.~Solano, and M.~Sanz.
\newblock Digital-analog quantum algorithm for the quantum {Fourier} transform.
\newblock {\em Physical Review Research}, 2(1):013012, January 2020.

\bibitem{celeri_digital-analog_2021}
L.~C. Celeri, D.~Huerga, F.~Albarrán-Arriagada, E.~Solano, and M.~Sanz.
\newblock Digital-analog quantum simulation of fermionic models.
\newblock {\em arXiv:2103.15689 [quant-ph]}, March 2021.
\newblock arXiv: 2103.15689.

\bibitem{lamata_digital-analog_2017}
L.~Lamata.
\newblock Digital-analog quantum simulation of generalized {Dicke} models with
  superconducting circuits.
\newblock {\em Scientific Reports}, 7(1):43768, March 2017.

\bibitem{arrazola_digital-analog_2016}
I.~Arrazola, J.~S. Pedernales, L.~Lamata, and E.~Solano.
\newblock Digital-{Analog} {Quantum} {Simulation} of {Spin} {Models} in
  {Trapped} {Ions}.
\newblock {\em Scientific Reports}, 6(1):30534, July 2016.

\bibitem{galicia_enhanced_2020}
A.~Galicia, B.~Ramon, E.~Solano, and M.~Sanz.
\newblock Enhanced connectivity of quantum hardware with digital-analog
  control.
\newblock {\em Physical Review Research}, 2(3):033103, July 2020.

\bibitem{babukhin_hybrid_2020}
D.~V. Babukhin, A.~A. Zhukov, and W.~V. Pogosov.
\newblock Hybrid digital-analog simulation of many-body dynamics with
  superconducting qubits.
\newblock {\em Physical Review A}, 101(5):052337, May 2020.

\bibitem{kairys_parametrized_2021}
Paul Kairys and Travis~S. Humble.
\newblock Parametrized hamiltonian simulation using quantum optimal control.
\newblock {\em Phys. Rev. A}, 104:042602, Oct 2021.

\bibitem{glaser_training_2015}
Steffen~J. Glaser, Ugo Boscain, Tommaso Calarco, Christiane~P. Koch, Walter
  Köckenberger, Ronnie Kosloff, Ilya Kuprov, Burkhard Luy, Sophie Schirmer,
  Thomas Schulte-Herbrüggen, Dominique Sugny, and Frank~K. Wilhelm.
\newblock Training {Schrödinger}’s cat: quantum optimal control: {Strategic}
  report on current status, visions and goals for research in {Europe}.
\newblock {\em The European Physical Journal D}, 69(12):279, December 2015.

\bibitem{fowler_surface_2012}
Austin~G. Fowler, Matteo Mariantoni, John~M. Martinis, and Andrew~N. Cleland.
\newblock Surface codes: Towards practical large-scale quantum computation.
\newblock {\em Phys. Rev. A}, 86:032324, Sep 2012.

\bibitem{terhal2015quantum}
Barbara~M Terhal.
\newblock Quantum error correction for quantum memories.
\newblock {\em Reviews of Modern Physics}, 87(2):307, 2015.

\bibitem{senthil_symmetry-protected_2015}
T.~Senthil.
\newblock Symmetry-{Protected} {Topological} {Phases} of {Quantum} {Matter}.
\newblock {\em Annual Review of Condensed Matter Physics}, 6(1):299--324, 2015.

\bibitem{affleck_rigorous_1987}
Ian Affleck, Tom Kennedy, Elliott~H. Lieb, and Hal Tasaki.
\newblock Rigorous results on valence-bond ground states in antiferromagnets.
\newblock {\em Physical Review Letters}, 59(7):799--802, August 1987.

\bibitem{pollmann_entanglement_2010}
Frank Pollmann, Ari~M. Turner, Erez Berg, and Masaki Oshikawa.
\newblock Entanglement spectrum of a topological phase in one dimension.
\newblock {\em Physical Review B}, 81(6):064439, February 2010.

\bibitem{pollmann_detection_2012}
Frank Pollmann and Ari~M. Turner.
\newblock Detection of symmetry-protected topological phases in one dimension.
\newblock {\em Physical Review B}, 86(12):125441, September 2012.

\bibitem{calvanese_strinati_destruction_2016}
Marcello Calvanese~Strinati, Leonardo Mazza, Manuel Endres, Davide Rossini, and
  Rosario Fazio.
\newblock Destruction of string order after a quantum quench.
\newblock {\em Physical Review B}, 94(2):024302, July 2016.

\bibitem{arute_quantum_2019}
F.~Arute, K.~Arya, R.~Babbush, D.~Bacon, J.~C. Bardin, R.~Barends, R.~Biswas,
  S.~Boixo, F.~GSL. Brandao, D.~A. Buell, et~al.
\newblock Quantum supremacy using a programmable superconducting processor.
\newblock {\em Nature}, 574(7779):505--510, October 2019.

\bibitem{blok2021quantum}
MS~Blok, VV~Ramasesh, T~Schuster, K~O’Brien, JM~Kreikebaum, D~Dahlen,
  A~Morvan, Beni Yoshida, NY~Yao, and I~Siddiqi.
\newblock Quantum information scrambling on a superconducting qutrit processor.
\newblock {\em Physical Review X}, 11(2):021010, 2021.

\bibitem{mckay2018qiskit}
David~C McKay, Thomas Alexander, Luciano Bello, Michael~J Biercuk, Lev Bishop,
  Jiayin Chen, Jerry~M Chow, Antonio~D C{\'o}rcoles, Daniel Egger, Stefan
  Filipp, et~al.
\newblock Qiskit backend specifications for openqasm and openpulse experiments.
\newblock {\em arXiv preprint arXiv:1809.03452}, 2018.

\bibitem{affleck1988valence}
Ian Affleck, Tom Kennedy, Elliott~H Lieb, and Hal Tasaki.
\newblock Valence bond ground states in isotropic quantum antiferromagnets.
\newblock In {\em Condensed matter physics and exactly soluble models}, pages
  253--304. Springer, 1988.

\bibitem{ueda2008finite}
Hiroshi Ueda, Hiroki Nakano, and Koichi Kusakabe.
\newblock Finite-size scaling of string order parameters characterizing the
  haldane phase.
\newblock {\em Physical Review B}, 78(22):224402, 2008.

\bibitem{yip2003dimer}
SK~Yip.
\newblock Dimer state of spin-1 bosons in an optical lattice.
\newblock {\em Physical review letters}, 90(25):250402, 2003.

\bibitem{garcia2004implementation}
Juan~J Garcia-Ripoll, Miguel~A Martin-Delgado, and J~Ignacio Cirac.
\newblock Implementation of spin hamiltonians in optical lattices.
\newblock {\em Physical review letters}, 93(25):250405, 2004.

\bibitem{imambekov2003spin}
Adilet Imambekov, Mikhail Lukin, and Eugene Demler.
\newblock Spin-exchange interactions of spin-one bosons in optical lattices:
  Singlet, nematic, and dimerized phases.
\newblock {\em Physical Review A}, 68(6):063602, 2003.

\bibitem{hilker2017revealing}
Timon~A Hilker, Guillaume Salomon, Fabian Grusdt, Ahmed Omran, Martin Boll,
  Eugene Demler, Immanuel Bloch, and Christian Gross.
\newblock Revealing hidden antiferromagnetic correlations in doped hubbard
  chains via string correlators.
\newblock {\em Science}, 357(6350):484--487, 2017.

\bibitem{cohen2014proposal}
Itsik Cohen and Alex Retzker.
\newblock Proposal for verification of the haldane phase using trapped ions.
\newblock {\em Physical review letters}, 112(4):040503, 2014.

\bibitem{cohen2015simulating}
I~Cohen, P~Richerme, Z-X Gong, C~Monroe, and A~Retzker.
\newblock Simulating the haldane phase in trapped-ion spins using optical
  fields.
\newblock {\em Physical Review A}, 92(1):012334, 2015.

\bibitem{senko2015realization}
C~Senko, P~Richerme, J~Smith, A~Lee, I~Cohen, A~Retzker, and C~Monroe.
\newblock Realization of a quantum integer-spin chain with controllable
  interactions.
\newblock {\em Physical Review X}, 5(2):021026, 2015.

\bibitem{albarran2018spin}
F~Albarr{\'a}n-Arriagada, L~Lamata, E~Solano, G~Romero, and JC~Retamal.
\newblock Spin-1 models in the ultrastrong-coupling regime of circuit qed.
\newblock {\em Physical Review A}, 97(2):022306, 2018.

\bibitem{krantz_quantum_2019}
P.~Krantz, M.~Kjaergaard, F.~Yan, T.~P. Orlando, S.~Gustavsson, and W.~D.
  Oliver.
\newblock A quantum engineer's guide to superconducting qubits.
\newblock {\em Applied Physics Reviews}, 6(2):021318, June 2019.

\bibitem{wu_high-fidelity_2020}
Xian Wu, S.~L. Tomarken, N.~Anders Petersson, L.~A. Martinez, Yaniv~J. Rosen,
  and Jonathan~L. DuBois.
\newblock High-{Fidelity} {Software}-{Defined} {Quantum} {Logic} on a
  {Superconducting} {Qudit}.
\newblock {\em Physical Review Letters}, 125(17):170502, October 2020.

\bibitem{li2022realization}
Shaowei Li, Daojin Fan, Ming Gong, Yangsen Ye, Xiawei Chen, Yulin Wu, Huijie
  Guan, Hui Deng, Hao Rong, He-Liang Huang, et~al.
\newblock Realization of fast all-microwave controlled-z gates with a tunable
  coupler.
\newblock {\em Chinese Physics Letters}, 39(3):030302, 2022.

\bibitem{yan_tunable_2018}
F.~Yan, P.~Krantz, Y.~Sung, M.~Kjaergaard, D.~L. Campbell, T.~P. Orlando,
  S.~Gustavsson, and W.~D. Oliver.
\newblock Tunable {Coupling} {Scheme} for {Implementing} {High}-{Fidelity}
  {Two}-{Qubit} {Gates}.
\newblock {\em Physical Review Applied}, 10(5):054062, November 2018.

\bibitem{sung_realization_2020}
Y.~Sung, L.~Ding, J.~Braum\"uller, A.~Veps\"al\"ainen, B.~Kannan,
  M.~Kjaergaard, A.~Greene, G.~O. Samach, C.~McNally, D.~Kim, et~al.
\newblock Realization of high-fidelity cz and $zz$-free iswap gates with a
  tunable coupler.
\newblock {\em Phys. Rev. X}, 11:021058, Jun 2021.

\bibitem{riviello_searching_2015}
Gregory Riviello, Katharine~Moore Tibbetts, Constantin Brif, Ruixing Long,
  Re-Bing Wu, Tak-San Ho, and Herschel Rabitz.
\newblock Searching for quantum optimal controls under severe constraints.
\newblock {\em Physical Review A}, 91(4):043401, April 2015.

\bibitem{egger_adaptive_2014}
D.~J. Egger and F.~K. Wilhelm.
\newblock Adaptive {Hybrid} {Optimal} {Quantum} {Control} for {Imprecisely}
  {Characterized} {Systems}.
\newblock {\em Physical Review Letters}, 112(24):240503, June 2014.

\bibitem{knee2015optimal}
George~C Knee and William~J Munro.
\newblock Optimal trotterization in universal quantum simulators under faulty
  control.
\newblock {\em Physical Review A}, 91(5):052327, 2015.

\bibitem{caneva_optimal_2009}
T.~Caneva, M.~Murphy, T.~Calarco, R.~Fazio, S.~Montangero, V.~Giovannetti, and
  G.~E. Santoro.
\newblock Optimal {Control} at the {Quantum} {Speed} {Limit}.
\newblock {\em Physical Review Letters}, 103(24):240501, December 2009.

\bibitem{kirchhoff_optimized_2018}
S.~Kirchhoff, T.~Kebler, P.~J. Liebermann, E.~Assémat, S.~Machnes, F.~Motzoi,
  and F.~K. Wilhelm.
\newblock Optimized cross-resonance gate for coupled transmon systems.
\newblock {\em Physical Review A}, 97(4):042348, April 2018.

\bibitem{lysne2020small}
Nathan~K Lysne, Kevin~W Kuper, Pablo~M Poggi, Ivan~H Deutsch, and Poul~S
  Jessen.
\newblock Small, highly accurate quantum processor for intermediate-depth
  quantum simulations.
\newblock {\em Physical review letters}, 124(23):230501, 2020.

\bibitem{barends_digitized_2016}
R.~Barends, A.~Shabani, L.~Lamata, J.~Kelly, A.~Mezzacapo, U.~Las Heras,
  R.~Babbush, A.~G. Fowler, B.~Campbell, Yu~Chen, Z.~Chen, B.~Chiaro,
  A.~Dunsworth, E.~Jeffrey, E.~Lucero, A.~Megrant, J.~Y. Mutus, M.~Neeley,
  C.~Neill, P.~J.~J. O'Malley, C.~Quintana, P.~Roushan, D.~Sank,
  A.~Vainsencher, J.~Wenner, T.~C. White, E.~Solano, H.~Neven, and John~M.
  Martinis.
\newblock Digitized adiabatic quantum computing with a superconducting circuit.
\newblock {\em Nature}, 534(7606):222--226, June 2016.
\newblock arXiv: 1511.03316.

\bibitem{kandala_hardware-efficient_2017}
Abhinav Kandala, Antonio Mezzacapo, Kristan Temme, Maika Takita, Markus Brink,
  Jerry~M. Chow, and Jay~M. Gambetta.
\newblock Hardware-efficient variational quantum eigensolver for small
  molecules and quantum magnets.
\newblock {\em Nature}, 549(7671):242--246, September 2017.

\bibitem{kraus2008preparation}
Barbara Kraus, Hans~P B{\"u}chler, Sebastian Diehl, Adrian Kantian, Andrea
  Micheli, and Peter Zoller.
\newblock Preparation of entangled states by quantum markov processes.
\newblock {\em Physical Review A}, 78(4):042307, 2008.

\bibitem{verstraete2009quantum}
Frank Verstraete, Michael~M Wolf, and J~Ignacio Cirac.
\newblock Quantum computation and quantum-state engineering driven by
  dissipation.
\newblock {\em Nature physics}, 5(9):633--636, 2009.

\bibitem{kairys2021efficient}
Paul Kairys and Travis~S Humble.
\newblock Efficient quantum gate discovery with optimal control.
\newblock In {\em 2021 IEEE International Conference on Quantum Computing and
  Engineering (QCE)}, pages 413--418. IEEE, 2021.

\bibitem{machnes_tunable_2018}
S.~Machnes, E.~Assémat, D.~Tannor, and F.~K. Wilhelm.
\newblock Tunable, {Flexible}, and {Efficient} {Optimization} of {Control}
  {Pulses} for {Practical} {Qubits}.
\newblock {\em Physical Review Letters}, 120(15):150401, April 2018.

\bibitem{palao_optimal_2003}
J.~P. Palao and R.~Kosloff.
\newblock Optimal control theory for unitary transformations.
\newblock {\em Physical Review A}, 68(6):062308, December 2003.

\bibitem{tripathi_operation_2019}
Vinay Tripathi, Mostafa Khezri, and Alexander~N Korotkov.
\newblock Operation and intrinsic error budget of a two-qubit cross-resonance
  gate.
\newblock {\em Physical Review A}, 100(1):012301, 2019.

\bibitem{Bezanson_Julia_A_fresh_2017}
Jeff Bezanson, Alan Edelman, Stefan Karpinski, and Viral~B. Shah.
\newblock {Julia: A fresh approach to numerical computing}.
\newblock {\em SIAM Review}, 59(1):65--98, 9 2017.

\bibitem{rackauckas2017differentialequations}
C.~Rackauckas and Q.~Nie.
\newblock Differentialequations.jl--a performant and feature-rich ecosystem for
  solving differential equations in julia.
\newblock {\em Journal of Open Research Software}, 5(1), 2017.

\bibitem{mogensen2018optim}
P.~K. Mogensen and A.N. Riseth.
\newblock Optim: A mathematical optimization package for {Julia}.
\newblock {\em Journal of Open Source Software}, 3(24):615, 2018.

\end{thebibliography}

\end{document}